\documentclass[12pt]{article}

\usepackage{setspace}
\usepackage[dviwin]{graphicx}
\usepackage{amsfonts}
\usepackage{amssymb}
\usepackage{mathrsfs}
\usepackage{amsmath}
\usepackage{epsfig}
\usepackage{relsize}
\usepackage{graphicx}
\usepackage{mathabx}

\usepackage{dsfont}
\usepackage{mathrsfs}
\usepackage{esint}
\usepackage{textpos}
\usepackage{wrapfig}
\usepackage{lipsum}
\usepackage{hyperref}
\hypersetup{
    colorlinks = true,
    linkcolor = [rgb]{0 .4 .8},
    citecolor=[rgb]{1 .2 .2}
}
\usepackage{array}

\newcommand{\bea}{\begin{equation}}
\newcommand{\eea}{\end{equation}}
\newcommand{\bear}{\begin{eqnarray}}
\newcommand{\eear}{\end{eqnarray}}
\newcommand{\bearr}{\begin{eqnarray*}}
\newcommand{\eearr}{\end{eqnarray*}}

\newcommand{\beal}{\begin{align}}
\newcommand{\eeal}{\end{align}}
\newcommand{\beall}{\begin{align*}}
\newcommand{\eeall}{\end{align*}}

\newcommand{\CP}{\mathds{C}\mathds{P}}
\newcommand{\CC}{\mathds{C}}

\newcommand{\dd}{\partial}

\newcommand{\comment}[1]{}

\newcommand{\dP}{\mathbf{dP}}

\usepackage{xcolor}
\usepackage{empheq}

\makeatletter
\def\@seccntformat#1{\@ifundefined{#1@cntformat}%
{\csname the#1\endcsname\quad}% default
{\csname #1@cntformat\endcsname}% individual control
}
\def\section@cntformat{{\normalfont\large\thesection.}\quad}
\def\subsection@cntformat{\textsection\, \thesubsection.\quad}
\def\subsubsection@cntformat{\textsection\textsection\, \thesubsubsection.\quad}
\makeatother

\newcommand{\ssection}[1]{%
  \section[#1]{\centering\normalfont\scshape #1}}
\newcommand{\ssubsection}[1]{%
   \subsection[#1]{\raggedright\normalfont  #1}}
\newcommand{\ssubsubsection}[1]{%
   \subsubsection[#1]{\raggedright\normalfont  #1}}

\newsavebox\MBox

\begin{document}

\title{Comments on the del Pezzo cone}
\author{Dmitri Bykov\footnote{Emails:
dmitri.bykov@aei.mpg.de, dbykov@mi.ras.ru}  \\ \\
{\small $\bullet$ Max-Planck-Institut f\"ur Gravitationsphysik, Albert-Einstein-Institut,} \\ {\small Am M\"uhlenberg 1, D-14476 Potsdam-Golm, Germany} \\ {\small $\bullet$ Steklov
Mathematical Institute of Russ. Acad. Sci., Gubkina str. 8, 119991 Moscow, Russia \;}}
%\date{18.05.2013 / ver. 27.06.2013}
\date{}
\maketitle
\vspace{-0.5cm}
\begin{center}
\line(1,0){450}
\end{center}
\vspace{-0.3cm}
\textbf{Abstract.} We describe a framework for constructing the general Ricci-flat metric on the anticanonical cone over the del Pezzo surface of rank one.
\vspace{-0.4cm}
\begin{center}
\line(1,0){450}
\end{center}

%\begin{textblock}{4}(9,-6.5)
\begin{textblock}{4}(9,-5.8)
\underline{\emph{AEI-2014-014}}
\end{textblock}

Whereas Ricci-flat metrics on \emph{compact} Calabi-Yau manifolds are difficult to construct, there exist many explicitly known Ricci-flat metrics on \emph{noncompact} Calabi-Yau manifolds (the first examples being \cite{EH}, \cite{GH}, \cite{CdO}). The reason is that these latter metrics possess sufficiently many isometries. The role of these metrics is that they describe the geometry of the compact Calabi-Yau manifold in the vicinity of a singularity, after it has been resolved. One particular type of singularity that can occur for a complex Calabi-Yau threefold is that of a cone over a complex surface. The goal of this article is to provide a framework for constructing the most general Ricci-flat metric (with the relevant isometries) on the anticanonical cone over the del Pezzo surface of rank one --- the blow-up of $\CP^2$ at one point. The metric of \cite{LuPope1}, which can be found by the so-called orthotoric ansatz of \cite{Gauduchon}, fits in our construction as a particular case.

\vspace{0.3cm}
The structure of the paper is as follows. In \S\,\ref{pezzo} we give definitions of the del Pezzo surface $\dP_1$, both an algebraic (in \S\, \ref{alg}) and a differential-geometric one (in \S\,\ref{diff}), and write out a Ricci-flatness equation for the metric on the anticanonical cone over $\dP_1$, that we set out to solve. In \S\, \ref{biangle} we introduce the moment polytope for a $U(1)^2$ action on the cone and describe its topological properties. In particular, we determine the normal bundles of the two $\CP^1$'s embedded in the corners of the polytope.

In \S\,\ref{expansionsec} we introduce our main technical tool -- the expansion of the metric at `infinity', i.e. away from the vertex of the cone. The leading order (\S\,\ref{leading}, \S\,\ref{metricinf}) of the expansion corresponds to a real cone over an Einstein-Sasaki manifold, whose only parameter is fixed by topological requirements -- the normal bundles of the spheres embedded in the del Pezzo surface, discussed previously. In \S \, \ref{regreq} we elaborate on a regularity requirement on the metric (or rather on a potential $G$ which determines the metric) near the edges of the moment polytope. This requirement has direct consequences for the coefficient functions of the expansion at `infinity'. In \S\,\ref{genericorder} we present the general structure of these coefficient functions. In \S\,\ref{Heunsec} we analyze the linear inhomogeneous equation arising in a generic order of perturbation theory. We show that the corresponding homogeneous equation has a solution compatible with the regularity requirement only in two orders of perturbation theory. This means that the Ricci-flat metric on the cone may depend on two parameters at most.

In \S\,\ref{orthometric} we demonstrate that one known metric on the cone over $\dP_1$ -- the so-called orthotoric one -- perfectly fits in our general considerations of the previous sections as a particular case, when the two potential parameters are related in a certain way. In \S\,\ref{orthodeform} we directly construct a first-order deformation of the orthotoric metric, which corresponds to the second parameter and has the asymptotic behavior at infinity predicted by our general analysis. Finally, in \S\,\ref{conclusion} we discuss the geometric interpretation of this new parameter.

Appendices \ref{solxi} and \ref{proof1} are technical and are referred to in the main text, whereas in appendix \ref{perturbfunc} we write out explicitly the polynomials that appear in the first few orders of the expansion of the potential $G$ at infinity.

\ssection{The del Pezzo surface and the cone: geometry}\label{pezzo}

We start with a definition of the del Pezzo surface -- the main hero of the constructions to follow. We will be interested in the del Pezzo surface of rank one (or, equivalently, of degree $8$), further denoted by $\dP_1$, which is in a sense the simplest algebraic surface after $\CP^2$ -- it is the blow-up of $\CP^2$ at one point. Since we will be mainly dealing with differential-geometric structures, we wish to give a definition of $\dP_1$ in topological terms amenable to differential-geometric analysis:

\vspace{0.3cm}
\textbf{Definition.} The del Pezzo surface $\dP_1$ is a compact simply-connected K\"ahler manifold of complex dimension 2, such that $H^2(\dP_1, \mathbb{Z})=\mathbb{Z}^2$, and the intersection pairing on $H^2(\dP_1, \mathbb{Z})$ has the form $\left( \begin{array}{ccc}
1 & 0 \\
0 & -1 \end{array} \right)$.
\vspace{0.5cm}

\ssubsection{The algebraic model}\label{alg}

There is a concrete algebraic model for the del Pezzo surface of rank one. The surface can be embedded into $\CP^8$, and the embedding is given by those sections of $\mathcal{O}_{\CP^2}(3)$ which vanish at a given point on $\CP^2$, for example at $(z_1:z_2:z_3)=(0:0:1)$. We can choose a monomial basis for these sections of $\mathcal{O}_{\CP^2}(3)$:
\bear\label{dP1}
&& x_1=z_1^3,\;x_2=z_1^2 z_2,\;x_3=z_1^2 z_3,\;x_4=z_1 z_2^2,\;x_5=z_1 z_3^2,\;\\ \nonumber
&& x_6=z_1 z_2 z_3,\;x_7=z_2^2 z_3,\;x_8=z_2 z_3^2,\;x_9=z_2^3
\eear
These are all possible cubic monomials in 3 variables with $z_3^3$ omitted, since it is the only one that does not vanish at the prescribed point. If we now regard $(x_1, \ldots, x_9)$ as homogeneous coordinates on the projective space $\CP^8$, then the above formulas (\ref{dP1}) provide the embedding. This embedding is called \emph{anticanonical}, since the standard tautological sheaf $\mathcal{O}_{\CP^8}(1)$ over the ambient space $\CP^8$, when restricted to the surface, coincides with its anticanonical sheaf.

The variables $x_1 \ldots x_9$ are not independent, and they satisfy a wealth of equations (an overdetermined system, i.e. the one with syzygies), e.g. the following ones:
\bear\label{dP1eqs}
\begin{array}{ccc}
x_1 x_6=x_2 x_3, &\quad x_1 x_5=x_3^2, &\quad x_1 x_4=x_2^2,\;\quad\ldots
\end{array}
\eear
The metric induced by this embedding on $\dP_1$ from the canonical Fubini-Study metric on $\CP^8$ is a well-defined metric on the del Pezzo surface.

Once the embedding is specified, i.e. when the equations (\ref{dP1eqs}) are given, the \emph{affine cone} may be constructed simply by passing from projective space $\CP^8$ to the affine space $\CC^9$, i.e. by treating the same set of equations (\ref{dP1eqs}) as written in $\CC^9$. Clearly, this produces a singularity at the vertex of the cone, when $x_1=\ldots=x_9=0$. It can be subsequently resolved to produce a smooth algebraic variety.

\ssubsection{The differential-geometric model}\label{diff}

In the construction of the previous section we chose a reference point $(0: 0: 1) \in \CP^2$, which was subsequently blown-up. This reduces the automorphism group $PGL(3, \CC)$ of $\CP^2$ to the automorphism group of $\dP_1$:
\bea\label{aut}
Aut (\dP_1)= P \left(
\begin{array}{ccc}
\bullet & \bullet & 0  \\
\bullet & \bullet & 0  \\
\bullet & \bullet & \bullet \\
\end{array}
 \right),
\eea
hence the cone $Y:=\mathrm{Cone}(\dP_1)$ has as its automorphism group the maximal parabolic subgroup of $GL(3, \CC)$ defined by matrices of the form (\ref{aut}) (forgetting the projectivization).

We will be looking for a K\"ahler metric on $Y$ with the isometry group being the maximal compact subgroup of $Aut (Y)$:
\bea
\mathrm{Isom}(Y)=U(2)\times U(1)
\eea
In more practical terms, we will introduce three complex coordinates $z_1, z_2, u$ on $Y$  and, due to the $U(2) \times U(1)$ isometry, we will assume that the K\"ahler potential depends on the two combinations of them:
\bea\label{Kahpot}
K=K(|z_1|^2+|z_2|^2, |u|^2)
\eea
The corresponding K\"ahler form is $\Omega= i \dd \bar{\dd} K$ and the metric is $g_{i \bar{j}}=\dd_i\bar{\dd}_j K$.
Since the Ricci tensor is related to the metric of a K\"ahler manifold as $R_{i\bar{j}}=-\dd_i\bar{\dd}_j \log \det g$, the Ricci-flatness (Calabi-Yau) condition $R_{i\bar{j}}=0$ implies that the determinant of the Hermitian metric $g$ has to factorize in a holomorphic and conjugate antiholomorphic pieces: $\det g=|f(z)|^2$. As $\det{g}$ is $U(2)\times U(1)$-invariant, it means that $\det g=a\,|u|^{2l}$ for some constants $a, l$. On the other hand, a direct calculation of $\det{g}$ for a metric arising from the K\"ahler potential (\ref{Kahpot}) gives
\bea
\det{g}=e^{-2t-s}\;K_t\,\left( K_{tt}K_{ss}-K_{ts}^2\right),
\eea
where
\bea\label{tsvars}
e^t=|z_1|^2+|z_2|^2\quad \mathrm{and}\quad e^s=|u|^2.
\eea
The Ricci-flatness condition is reduced to the following equation:
\bea\label{RicciflatK}
K_t\,\left( K_{tt}K_{ss}-K_{ts}^2\right)=a\,e^{2t+(l+1)\,s}
\eea
It turns out useful to perform a Legendre transform, passing from the variables $\{t, s\}$ to the new independent variables
\bea\label{munu}
\mu={\dd K \over \dd t},\quad\quad\nu={\dd K \over \dd s}
\eea
and from the K\"ahler potential $K(t, s)$ to the dual potential $G(\mu, \nu)$:
\bea
G=\mu\,t+\nu\,s-K
\eea
The usefulness of the new variables (\ref{munu}) to a large extent relies on the fact that they have a transparent geometric meaning -- these are the moment maps for the following two $U(1)$ actions on $Y$:
\bea\label{U1action}
U(1)_\mu:\quad (\,z_1\to e^{i\alpha}\,z_1,\quad z_2 \to e^{i\alpha}\,z_2\,)\quad\quad\quad
U(1)_\nu:\quad u \to e^{i\beta} \, u
\eea

In this paper we will leave aside the case $l+1=0$ ($l$ is the parameter entering the exponent in (\ref{RicciflatK})) and assume that $l+1\neq 0$. In this case we can get rid of the $l$ dependence by a rescaling $\nu \to (l+1)\,\nu$, so in what follows we effectively set $l=0$. Making one more rescaling $\mu \to 2\,\mu$, we obtain from (\ref{RicciflatK}) a Monge-Ampere equation for the dual potential $G$ -- a function of two variables $\mu, \nu$ -- of the following form:
\begin{empheq}[box=\fbox]{align}
\hspace{1em}\vspace{1em}
\label{Ricciflat}
e^{\frac{\dd G}{\dd \mu}+\frac{\dd G}{\dd \nu}}\;\left(\frac{\dd^2 G}{\dd \mu^2} \frac{\dd^2 G}{\dd \nu^2}-\left(\frac{\dd^2 G}{\dd \mu \dd\nu}\right)^2\right)=\tilde{a}\,\mu
\hspace{1em}
\end{empheq}

Denoting $(\mu, \nu)$ by $(\mu_1, \mu_2)$, we can recover the metric from the dual potential $G$ \cite{Pedersen} using the formula 
\bea\label{metric}
ds^2=\mu\, g_{\CP^1}+\sum\limits_{i, j=1}^2\,\frac{\dd^2 G}{\dd \mu_i \dd \mu_j}\,d\mu_i\,d\mu_j+\sum\limits_{i, j=1}^2\, \left(\frac{\dd^2 G}{\dd \mu^2}\right)^{-1}_{ij}\,\left(d\phi_i - A_i \right)\,\left(d \phi_j - A_j \right),
\eea
where $g_{\CP^1}$ is the standard round metric on $\CP^1$, $A_2=0$ and $A_1$ is the `K\"ahler current' of $\CP^1$, i.e. a connection, whose curvature is the Fubini-Study form of $\CP^1$: $d A_1=\omega_{\CP^1}$.

\vspace{0.3cm}
\underline{\emph{Comment.}} Note that the parameter $\tilde{a}$ in (\ref{Ricciflat}) is irrelevant, since one can effectively set $\tilde{a}=1$ by a \emph{linear} redefinition of the potential $G$, i.e. $G \to G+\nu \log{(\tilde{a})}$. Such a linear redefinition does not affect the metric (\ref{metric}), which depends only on the second derivatives of $G$.

\ssubsection{The moment `biangle'}\label{biangle}

\begin{figure}[h]
    \centering
    \includegraphics[width=\textwidth]{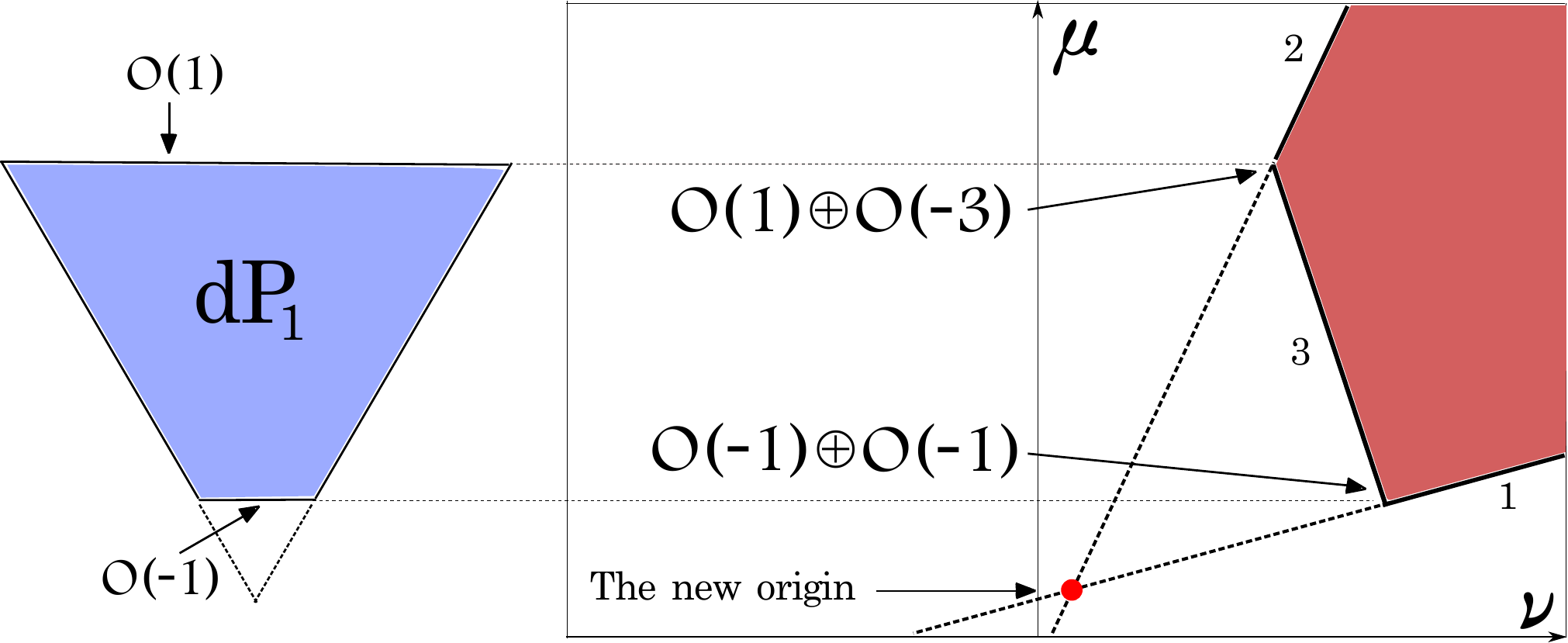}
    \caption{The trapezia -- the moment polygon of $\dP_1$ -- and the $(\mu, \nu)$ plane section of the moment polytope for the cone over $\dP_1$.}
    \label{mompol}
\end{figure}

Since $(\mu, \nu)$ are moment maps for the $U(1)^2$ action, the domain on which the potential $G(\mu, \nu)$ is defined is the moment polygon for this $U(1)^2$ action. In this case it is an unbounded domain with two vertices. Hence we may call it a `biangle', and it is depicted in Fig. \ref{mompol}.

From the perspective of the equation (\ref{Ricciflat}), it is the singularities of the function $G$ that determine the polytope. It is known 
\cite{Guillemin} that in the simplest case of a (generally non-Ricci-flat) metric induced by a K\"ahler quotient of flat space with respect to an action of a complex torus, the potential $G$ takes the form of a superposition of `hyperplanes':
\bea\label{Guilleminform}
G_{\mathrm{toric}}=\sum\limits_{i=1}^M\,L_i \,(\log{L_i}-1)\quad\mathrm{with}\quad L_i=\alpha_i \mu+\beta_i \nu+\gamma_i\, .
\eea
In general, a potential $G$ satisfying (\ref{Ricciflat}) will not have this form. However, we will assume that it has the corresponding \emph{asymptotic} behavior at the faces of the moment polytope. More exactly, when we approach an arbitrary face $L_i$, i.e. when $L_i \to 0$, we impose the asymptotic condition
\bea\label{asymptcond}
G=L_i \,(\log{L_i}-1)+\ldots\quad\mathrm{as}\quad L_i \to 0,
\eea
where the ellipsis indicates terms regular at $L_i \to 0$. Despite being subleading, they are important for the equation (\ref{Ricciflat}) to be consistent even in the limit $L_i \to 0$.  Consistency of the equation as well requires that $\alpha_i+\beta_i=1$.

Notice that, in addition to the $U(1)^2$ action (\ref{U1action}), there is yet another $U(1)$ action given by $(z_1\to e^{i\alpha}\,z_1,\; z_2 \to e^{-i\alpha}\,z_2)$, so Fig. \ref{mompol} corresponds in fact to a section of a three-dimensional moment polytope. Therefore the fiber over a generic point of this section -- a point in the interior -- is $\CP^1 \times \mathbb{T}^2$. The third $U(1)$ action corresponds to the rotation of the sphere $\CP^1$ around its axis. We will now demonstrate how the angles of the moment polytope are detemined by the normal bundles to the two $\CP^1$'s `located' in the corners.

A corner of the moment polytope may be given by the equations
\bea
\lambda_1=0,\quad \lambda_2=0,
\eea
where
\bea\label{lines}
\lambda_i=\alpha_i \mu+ \beta_i \nu+\gamma_i,\quad i=1, 2
\eea
are two linear forms. Moreover, according to the discussion above we assume that the behavior of the potential $G$ near the corner is as follows:
\bea\label{Gcorner}
G=\lambda_1 (\log{\lambda_1}-1)+\lambda_2 (\log{\lambda_2}-1)+\ldots,
\eea
where $\ldots$ denotes less singular terms. Compatibility with the Ricci-flatness condition (\ref{Ricciflat}) implies
\bea
\alpha_i+\beta_i=1,\quad i=1, 2 
\eea

We wish to determine what the behavior (\ref{Gcorner}) implies for the \emph{metric} near a given embedded $\CP^1$. For this purpose we perform a Legendre transform, passing from $\{\mu, \nu\}$ back to the dual variables $\{t, s\}$ (see (\ref{tsvars})) and calculating the K\"ahler potential
\bea
K=\kappa\;\log{\left(|z_1|^2+|z_2|^2\right)}+\left( |z_1|^2+|z_2|^2 \right)^n\;|u|^{n'}+\left( |z_1|^2+|z_2|^2 \right)^m\;|u|^{m'}+\ldots,
\eea
where $\kappa=2\,\frac{\gamma_2 \beta_1-\gamma_1 \beta_2}{\beta_2-\beta_1}$ and
\begin{empheq}[box=\fbox]{align}
\hspace{1em}\vspace{1em}
\label{normalbundle}
n={2\beta_2 \over \beta_2-\beta_1},\quad m=-{2 \beta_1 \over \beta_2-\beta_1},
\hspace{1em}
\end{empheq}
$$
n'=-{2(1-\beta_2)\over \beta_2-\beta_1},\quad m'={2(1-\beta_1)\over \beta_2 -\beta_1}
$$
Upon changing the complex coordinates according to the rule
\bea
w:={z_2 \over z_1},\quad x:=z_1^n\,u^{n'},\quad y:=z_1^m\, u^{m'}
\eea
we can bring the K\"ahler potential to the form
\begin{empheq}[box=\fbox]{align}
\hspace{1em}
\label{Kahlercorner}
\renewcommand{\arraystretch}{2}
\begin{tabular}{m{0.7 \textwidth}}
\vspace{-0.6cm}
$K=\kappa\;\log{\left(1+|w|^2\right)}+\left( 1+|w|^2 \right)^n\;|x|^2+\left( 1+|w|^2 \right)^m\;|y|^2+\ldots,$
\end{tabular}
\vspace{1em}\hspace{1em}
\end{empheq}
For $\kappa >0$ this implies that the normal bundle $N_{\CP^1}$ to the $\CP^1$ parametrized by the inhomogeneous coordinate $w$ and located in a given corner of the moment polytope is\footnote{See \cite{Bykov} for a detailed discussion of how the K\"ahler potential encodes the normal bundle to a $\CP^1$ in the analogous situation, when the $\CP^1$ is embedded in a complex surface.}
\bea
N_{\CP^1}=\mathcal{O}(-n)\;\oplus\;\mathcal{O}(-m),\quad\quad n+m=2
\eea
Note that $n+m=2$, as required by the Calabi-Yau condition
\bea
\det{N_{\CP^1}}=\;\textrm{the canonical class of}\;\CP^1\;=\mathcal{O}(-2)
\eea

In the del Pezzo cone case the two corners of the moment biangle in the $(\mu, \nu)$-plane correspond to the two bases of the trapezia representing the moment polytope of the del Pezzo surface itself, which serves as the base of the cone. This is emphasized in Fig. \ref{mompol}.  These two bases of the trapezia correspond to the two $\CP^1$'s embedded in the del Pezzo surface:
\begin{itemize}
\item One $\CP^1$ is inherited from $\CP^2$, i.e. it is the standard embedding $\CP^1 \hookrightarrow \CP^2$, hence the normal bundle inside $\dP_1$ is $N=\mathcal{O}(1)$. This implies that the normal bundle inside the \emph{cone over} $\dP_1$ is $N=\mathcal{O}(1)\,\oplus\,\mathcal{O}(-3)$
\item The second $\CP^1$ is the exceptional divisor of the blow-up and is embedded with normal bundle $N=\mathcal{O}(-1)$. The normal bundle inside the \emph{cone over} $\dP_1$ is therefore $N=\mathcal{O}(-1)\,\oplus\,\mathcal{O}(-1)$. 
\end{itemize}
These two spheres generate the second homology group of the del Pezzo surface, and their intersection matrix is $\left( \begin{array}{ccc}
1 & 0 \\
0 & -1 \end{array} \right)$. In particular, the diagonal $\pm 1$ entries encode the normal bundles to the spheres.

\ssection{An expansion away from the vertex of the cone}\label{expansionsec}

To start the analysis of the equation (\ref{Ricciflat}) first of all we shift the origin along the $\mu$-axis by a constant $\mu_0$ in such a way that the new origin is located at the intersection point of the two outer lines of the moment `biangle'. The new origin is indicated by the red dot in Fig. \ref{mompol}.

We aim at building an expansion of the metric at `infinity', i.e. far from the `vertex'. For this purpose, instead of the $\{ \mu, \nu\}$ variables, we will use a `radial' variable $\nu$ and an angular variable $\xi$:
\bea\label{newvars}
\{\mu, \nu\} \to \left\{ \nu,\;\xi={\mu- \mu_0 \over \nu}\right\}
\eea
Then the equation (\ref{Ricciflat}) above may be rewritten as follows:
\bea\label{Ricciflatnew}
e^{{\dd G \over \dd \nu}-{\xi-1\over \nu} {\dd G \over \dd \xi}}\;\left[{\dd^2 G \over \dd \xi^2} {\dd^2 G \over \dd \nu^2}-\left({\dd^2 G \over \dd \xi \dd \nu}-{1\over \nu}{\dd G\over \dd \xi}\right)^2\right]=a\;\nu^3 \left( \xi+{\mu_0\over \nu}\right)
\eea
We propose the following expansion for the potential $G$ at $\nu \to \infty$ ($b$ is a constant):

\begin{empheq}[box=\fbox]{align}
\hspace{1em}\vspace{1em}
\label{Gexp1}
G=3 \nu (\log{\nu}-1)+\nu\,P_0(\xi)+b \log{\nu}+\sum\limits_{k=0}^\infty\;\nu^{-k}\;P_{k+1}(\xi)
\hspace{1em}
\end{empheq}

Substituting this expansion in the above equation, we obtain a `master' equation, which can then be expanded in powers of ${1\over \nu}$ and solved iteratively for the functions $P_k(\xi)$:
\begin{empheq}[box=\fbox]{align}
\hspace{0.5em}
 \label{master} \sum\limits_{k=0}^\infty \nu^{-k}\,P_k''(\xi)\;\times\;\left(3-{b\over \nu}+\sum\limits_{k=2}^\infty\,k(k-1)P_k(\xi) \nu^{-k}\right)-\left(\sum\limits_{k=1}^\infty\,k P_k'(\xi)\,\nu^{-k} \right)^2=\hspace{1em}\\ \nonumber =a\,\left( \xi + {\mu_0\over \nu}\right)\,e^{-{b\over \nu}+\sum\limits_{k=0}^\infty\,\left( (\xi-1)\,P_k'+(k-1)\,P_k\right)\,\nu^{-k}}
\hspace{1em}
\end{empheq}
\ssubsection{Leading order}\label{leading}
\vspace{0.2cm}

The first equation is obtained from (\ref{master}) in the limit $\nu \to \infty$:
\bea
P_0''={a\over 3}\,\xi\,e^{(\xi-1) P_0'-P_0}
\eea
and has the solution
\bea
P_0(\xi)=\log{\left( -{a \over 9}\right)}-\sum\limits_{i=0}^2\;\frac{\xi-\xi_i}{\xi_i-1}\;\log{(\xi-\xi_i)},
\eea
where $\xi_i$ are the roots of the polynomial
\bea\label{Qpolynomial}
Q(\xi)=\xi^3-{3\over 2} \xi^2+d,
\eea
and $d$ is a constant of integration, which plays a crucial geometric role that we will now reveal.

\begin{wrapfigure}{l}{0.45\textwidth}
  \centering
    \includegraphics[width=0.45\textwidth]{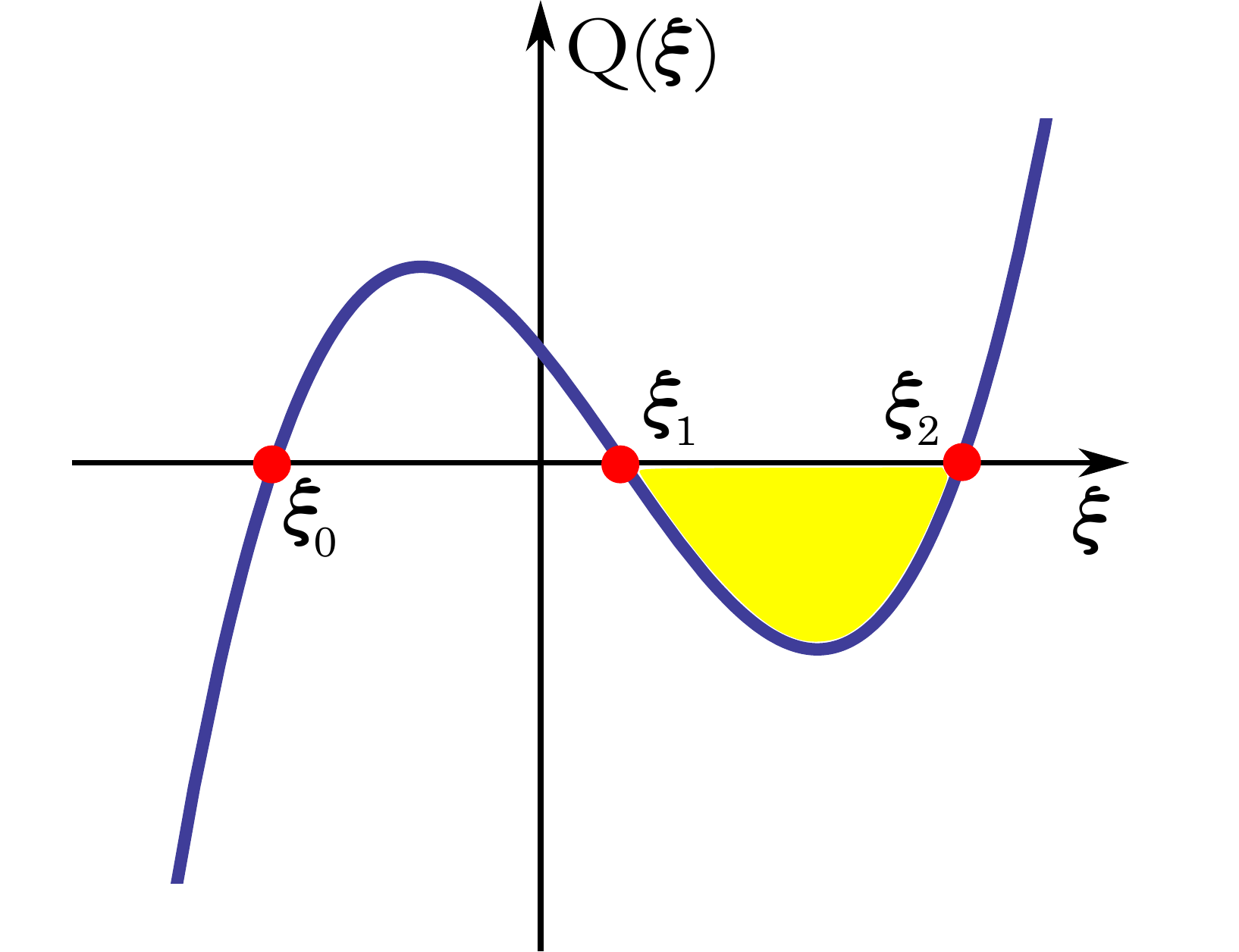}
    \caption{Yellow shading indicates the physical interval $\xi \in (\xi_1, \xi_2)$.}
    \label{Qxfunc}
\end{wrapfigure}

The singular case $\xi_1=1$ (and hence $\xi_2=1$) corresponds to the situation, when the physical region shown in Fig. \ref{Qxfunc} shrinks to zero (see next section). We will therefore omit it in our discussion.

\vspace{0.2cm}
\ssubsubsection{The metric at $\infty$}\label{metricinf}
\vspace{0.2cm}

The function $P_0(\xi)$ determines the metric at infinity by means of the formulas (\ref{Gexp1}) and (\ref{metric}). One can check that in the $(\nu, \xi)$ variables the `radial' part of the metric defined by
\bea\label{G0metr}
G_0=3 \,\nu (\log{\nu}-1)+\nu \,P_0(\xi)
\eea
looks as follows ($r=2 \,\sqrt{3 \nu}$):
\bea\label{infmetr}
\left[ds^2\right]_\mu:= \frac{\dd^2 G}{\dd \mu_i \dd \mu_j}\,d\mu_i d\mu_j=3 {d\nu^2 \over \nu}+\nu \,P_0''(\xi)\, d\xi^2=dr^2+r^2 \,{P_0'' \over 12}\,d\xi^2
\eea
In particular, we see that positivity of the metric requires $P_0''>0$.

The potential (\ref{G0metr}) may as well be written in the original $(\mu, \nu)$ variables (here we effectively set $\mu_0=0$):
\bea\label{G0}
G_0=\sum\limits_{i=0}^2\;\frac{\mu-\xi_i\,\nu}{1-\xi_i}\;\left(\log{(\mu-\xi_i\,\nu)}-1\right)
\eea
One sees that the slopes of the three lines involved are defined by the roots $\xi_i$:
\bea
\mathrm{Slope}_i=\left( {\mu \over \nu}\right)_i=\xi_i
\eea
It is important to mention that the three lines appearing in (\ref{G0}) are \emph{not} the three edges of the moment polytope depicted in Fig. \ref{mompol}. In fact, two of the lines, associated with the roots $\xi_1, \xi_2$, do correspond to the boundaries $1, 2$ of the polytope, however the line associated with the root $\xi_0$ is auxiliary and does not have a direct geometric interpretation.

In the notations (\ref{lines}) of the moment polytope, which we used before, one has
\bea
\xi_1=-\frac{\beta_1}{1-\beta_1}\quad \mathrm{and}\quad \xi_2=-\frac{\beta_2}{1-\beta_2}
\eea
On the other hand, $\beta_1$ and $\beta_2$ are both related to $\beta_3$ (the indices $1, 2, 3$ correspond to the numbering of lines in Fig. \ref{mompol}) through the normal bundle formulas (\ref{normalbundle}), which therefore implies that there is a relation between $\xi_1$ and $\xi_2$. This geometric relation fixes the parameter $d$ of the polynomial $Q(\xi)$.

Indeed, from the normal bundle formulas (\ref{normalbundle}) and Fig. \ref{mompol} it follows that
\bea
1-\frac{\beta_2}{\beta_3}=-2,\quad 1-\frac{\beta_1}{\beta_3}=2
\eea
Hence $\frac{\beta_2}{\beta_1}=-3$. This implies the following relation for $\xi_1, \xi_2$:
\bea\label{relxi}
-\frac{\xi_2}{1-\xi_2}=\frac{3 \xi_1}{1-\xi_1}
\eea

One can show (see Appendix \ref{solxi}) that it has two solutions: $(\xi_1^{(1)}, \xi_2^{(1)})$, $(\xi_1^{(2)}, \xi_2^{(2)})$. However, for $\xi \in (\xi_1^{(2)}, \xi_2^{(2)})$ one has $P_0''<0$ and for $\xi \in (\xi_1^{(1)}, \xi_2^{(1)})$ one has $P_0''>0$, so the positivity of the metric requires that we choose the first solution. It corresponds to
\bea
d=\frac{16+\sqrt{13}}{64}\,.
\eea
The third root of $Q(\xi)=0$, which we will denote $\xi_0$, is smaller than the two other roots (see Fig. \ref{Qxfunc}).

In what follows we will denote the roots of $Q(\xi)$ by $\xi_0, \xi_1, \xi_2$ so that $Q(\xi)=\prod\limits_{i=0}^2\;(\xi-\xi_i)$ and we will take into account that the `physical' region corresponds to $\xi \in (\xi_1, \xi_2)$.

\vspace{0.2cm}
\ssubsection{Regularity requirement}\label{regreq}
\vspace{0.2cm}

Recall that the edges of the moment polytope of Fig. \ref{mompol} are determined by the singularities of the potential $G(\mu, \nu)$. More exactly, we required that near each edge $L_i=0$ the function $G$ should behave as in (\ref{asymptcond}):
\bea
G=L_i \,(\log{L_i}-1)+\ldots\quad\mathrm{as}\quad L_i \to 0
\eea
By placing the origin at the intersection point of the lines $1, 2$ of Fig. \ref{mompol}, we make sure that the equations of these lines have the form
\bea
\mathrm{Line}\;1:\quad \mu-\mu_0=\xi_1 \,\nu,\quad\quad \mathrm{Line}\;2:\quad \mu-\mu_0=\xi_2 \,\nu,
\eea
to \emph{all orders} of perturbation theory. Indeed, the lines clearly cannot change their slopes, and $\xi_1, \xi_2$ are their slopes at infinity. The only thing that could happen in higher orders of perturbation theory is that the lines could shift and no longer pass through the origin $\mu=\nu=0$ (as they do in the 0-th order of perturbation theory, formula (\ref{G0})). As we consider the order $1\over \nu$ below, we will see that this does indeed happen, and precisely to account for this modification we shift the origin to the new intersection point of the two lines. To summarize, $G$ can be written as
\bea
G=\frac{\mu-\mu_0-\xi_1\,\nu}{1-\xi_1}\;\left(\log{(\mu-\mu_0-\xi_1\,\nu)}-1\right)+\frac{\mu-\mu_0-\xi_2\,\nu}{1-\xi_2}\;\left(\log{(\mu-\mu_0-\xi_2\,\nu)}-1\right)+\Delta,
\eea
where $\Delta$ is a function regular at $\mu-\mu_0=\xi_1 \nu$ and $\mu-\mu_0=\xi_2 \nu$. In terms of the $(\nu, \xi)$ variables the statement is that $\Delta(\nu, \xi)$ is regular at $\xi = \xi_1, \xi_2$ for any fixed $\nu$. In the forthcoming analysis of the higher orders of perturbation theory around infinity we will make the crucial \emph{assumption} that each term of the expansion of $\Delta(\nu, \xi)$ in powers of $1\over \nu$ is a function of $\xi$, regular at the two points $\xi = \xi_1, \xi_2$. We will see below that this requirement imposes extremely stringent conditions on the functions that can appear in higher orders of the perturbative expansion.

\vspace{0.2cm}
\ssubsection{Order $1\over \nu$}
\vspace{0.2cm}

The equation for $P_1(\xi)$, which arises as the coefficient of $\nu^{-1}$ in (\ref{master}), is as follows:
\bea
{d \over d \xi} \left( Q(\xi)\,{d P_1\over d \xi}\right)=2\,b\,\xi-3\,\mu_0
\eea
and leads to the solution
\bea
P_1=b\;\int\;{d \xi \over Q(\xi)}\;\left( \xi^2-{3 \mu_0\over b} \,\xi+\kappa\right),
\eea
where $\kappa$ is a constant of integration. If the numerator of the integrand is nonzero at a given root $\xi_1, \xi_2$ of $Q(\xi)$, $P_1(\xi)$ has a logarithmic singularity at this point, which is exactly what we wish to avoid by the regularity condition described in the previous section (a logarithmic singularity $\log{x}$ of $P_1$ would be more severe than the $  x \log{x}$ singularity of $P_0$). Therefore we will fix the constants $\kappa$ and ${\mu_0\over b}$ by the condition
\bea
 \xi^2-{3 \mu_0\over b} \,\xi+\kappa=(\xi-\xi_1)(\xi-\xi_2),
\eea
We see, in particular, that the first order of perturbation theory completely fixes the vertical shift of the lines by determining the constant $\mu_0$. The function $P_1$ then satisfies
\bea
P_1'(\xi)=\frac{b}{\xi-\xi_0}\, .
\eea
Note that, according to (\ref{Gexp1}), shifting the function $P_1(\xi)$ by a constant simply leads to a shift of the potential $G$ by a constant, and is therefore inessential.

\vspace{0.2cm}
\ssubsection{Arbitrary order}\label{genericorder}
\vspace{0.2cm}

It will be explained in the following sections that the function $G$ satisfying eq. (\ref{Ricciflatnew}) has the following structure:
\bear\label{Gexpgen}
G=3 \nu \left(\log{\nu}-1\right)+\nu\,P_0(\xi)+b \log{\left(\nu(\xi-\xi_0)\right)}+\sum\limits_{k=1}^\infty\;\nu^{-k}\;P_{k+1}(\xi)\\  \label{Gexpgen2}
\mathrm{with}\quad P_k(\xi)=b^k \left( \frac{(-1)^k}{k(k-1)}\,\left(\frac{1-\xi_0}{\xi-\xi_0}\right)^{k-1}+\mathrm{Polyn}_{k-3}(\xi)\right),\quad k\geq 2
\eear
As it should be clear from the notation, $\mathrm{Polyn}_{k-3}(\xi)$ is a polynomial of degree $k-3$ for $k\geq 3$ (and is zero for $k<3$). The polynomials that appear in the first few orders of the expansion are given explicitly in Appendix \ref{perturbfunc}.

The terms in (\ref{Gexpgen})-(\ref{Gexpgen2}) singular in $\xi-\xi_0$ can be easily summed to produce the following:
\bear\label{3line1}
G=\frac{\tilde{\mu}- \xi _1\,\nu}{1-\xi_1}
   \left(\log \left(\frac{\tilde{\mu}-  \xi _1\,\nu }{1-\xi _1}\right)-1\right)+\frac{\tilde{\mu}- \xi _2\,\nu}{1-\xi_2}
   \left(\log \left(\frac{\tilde{\mu}-  \xi _2\,\nu }{1-\xi _2}\right)-1\right)+\\ \nonumber+
\left(\frac{\tilde{\mu}- \xi _0\,\nu}{1-\xi_0}+b\right)\,\left(\log{\left(\frac{\tilde{\mu}- \xi _0\,\nu}{1-\xi_0}+b\right)}-1\right)+b\,\sum\limits_{k=2}^\infty\;\left(\frac{b}{\nu}\right)^k\;\mathrm{Polyn}_{k-2}(\xi)
\eear
Here the variable $\mu$ has been shifted in such a way that the new origin is located at $\tilde{\mu}=\nu=0$. The variable $\tilde{\mu}$ is therefore related to $\mu$ as $\tilde{\mu}=\mu-\mu_0$ with $\mu_0=b\,\frac{ \xi _1 \xi _2 \left(1-\xi _0\right)}{\left(\xi _1-\xi _0\right) \left(\xi _2-\xi _0\right)}$. Similarly, $\xi=\frac{\tilde{\mu}}{\nu}$.

Notice that the first three terms in (\ref{3line1}) have Guillemin's form (\ref{Guilleminform}). Interestingly, they provide an \emph{exact} solution of the equation (\ref{Ricciflat}) for an arbitrary value of $b$. Indeed, in the next sections we will show that all polynomials $\mathrm{Polyn}_{k}(\xi)$ depend on two parameters $\alpha, \beta$ and, in particular, the polynomials vanish for the zero values of these parameters. Therefore in the limit $\alpha=\beta=0$ the potential $G$ given by (\ref{3line1}) acquires Guillemin's form. It should be mentioned, however, that this `three-line' solution in general has `bad' singularities at the intersection points of the lines, and hence the underlying space is not a manifold. Nonetheless, for certain values of the parameter $d$, which determines the roots $\xi_i$ of the polynomial (\ref{Qpolynomial}), it provides a perfectly well-defined Einstein-Sasaki metric at `infinity'.

\vspace{0.2cm}
\ssubsection{Singular points of the Heun equation and eigenfunctions}\label{Heunsec}
\vspace{0.2cm}

We proceed to describe in more detail the equations that arise in higher orders of perturbation theory. Our goal is to explain the formula (\ref{3line1}) and elaborate on it.

In the $M$-th order of perturbation theory we arrive at the following equation:
\begin{empheq}[box=\fbox]{align}
\hspace{0.5em}\label{DMop}
D_M P_M := \frac{d}{d\xi}\left( Q(\xi) \frac{d P_M}{d\xi}\right)-\left((M-2)^2-1\right)\,\xi P_M=\textrm{r.h.s.},\hspace{1em}
\end{empheq}
where
\bea
Q(\xi)=\xi^3-{3\over 2} \xi^2+d
\eea
and the right hand side depends on the previous orders of perturbation theory, i.e. on $P_{M-1},\, \ldots,\, P_0$ and their derivatives. As discussed above, the del Pezzo cone corresponds to
\bea\label{dval}
d=\frac{16+\sqrt{13}}{64}.
\eea
As we claimed in (\ref{Gexpgen})-(\ref{Gexpgen2}), the inhomogeneous equation (\ref{DMop}) has a polynomial solution of degree $M-3$. This will be proven below, see \S\, \ref{expfuncpol}. The general solution, however, is produced by adding to this particular solution a general solution of the homogenized equation $D_M \Pi_M=0$.  The roots $\xi_i, i=0, 1, 2$ of the polynomial $Q(\xi)=\prod\limits_{i=0}^2\,(\xi-\xi_i)$ are singular points of this equation. Moreover, by making the change of variables $\xi\to {1\over \xi}$, one easily sees that $\infty$ is a singular point as well. Hence $D_M \Pi_M=0$ is a Fuchsian equation with 4 singular points -- a particular case of the so-called Heun equation, in which all exponents are zero.

The question we wish to pose is whether the homogenized equation $D_M \Pi_M=0$ has a nontrivial solution regular at \emph{two} of the singular points, say $\xi_1, \xi_2$. This is necessary in order to comply with the regularity requirement of \S\,\ref{regreq}. We claim that the answer is positive only for $M=3, 4$:
\bear
&&\Pi_3=\alpha\\ \label{betapar}
&&\Pi_4=\beta (\xi-1),\\ \nonumber
&&\textrm{where}\quad \alpha, \beta =\mathrm{const.}
\eear
Quite interestingly, the nontrivial solutions are independent of the constant $d$, which suggests that they are also relevant for the deformations of other Ricci-flat cones asymptotic to (real) cones over Sasaki-Einstein manifolds.

For what follows it will be convenient to parametrize the first two nonzero polynomials in (\ref{Gexpgen}) as follows:
\bea\label{P34}
\mathrm{Polyn}_0(\xi)=\alpha,\quad\quad \mathrm{Polyn}_1(\xi)= -{2\over 3}\,\alpha+\beta(\xi-1)
\eea
Here $\alpha$ and $\beta$ are the parameters of the metric.

In general the fact that some parameters are absent in the metric at infinity and then appear in different orders of the expansion around this metric is compatible with the known cases. One prominent example is the resolved conifold, which is asymptotic to the real cone over the Einstein-Sasaki manifold $T^{1,1}=\frac{SU(2)\times SU(2)}{U(1)}$ at infinity and exhibits two resolution parameters in an expansion around infinity \cite{PZTmain}.

\ssubsubsection{The proof}

We start with a proof that the equation $D_2 \Pi_2=\frac{d}{d\xi}\left(Q(\xi)\frac{d\Pi_2}{d\xi}\right)+\Pi_2=0$ has no nontrivial solution regular at $\xi=\xi_1, \xi_2$. Multiplying the equation by $\Pi_2$ and integrating over $\xi$ from $\xi_1$ to $\xi_2$, we obtain:
\bea
0=\int\limits_{\xi_1}^{\xi_2}\,d\xi\;\Pi_2\,\left( \frac{d}{d\xi}\left(Q(\xi)\frac{d\Pi_2}{d\xi}\right)+\Pi_2 \right)=
\int\limits_{\xi_1}^{\xi_2}\,d\xi\, \left( -Q(\xi)\,\left(\frac{d \Pi_2}{d\xi}\right)^2+\Pi_2^2\right),
\eea
where in the second equality we have integrated by parts. Since, according to Fig. \ref{Qxfunc}, $Q(\xi)<0$ for $\xi\in(\xi_1, \xi_2)$, we obtain $\Pi_2(\xi)=0$.

\vspace{0.5cm}The strategy of the proof that there are no solutions for $M\geq 5$ consists of the following two steps:
\begin{enumerate}
\item Proving that there are no polynomial solutions for $M \geq 5$
\item Assuming there is a (nonpolynomial) solution $P_M$ regular at two singular points, one can expand it in Legendre polynomials. By the analysis of the recurrence relation one can prove that the expansion is divergent.
\end{enumerate}

\vspace{0.3cm}
\textbf{Conjecture 1.} The homogeneous Heun equation $D_M \Pi_M=0$ has no \underline{\emph{polynomial}} solutions for $M\geq 5$ and $d$ given by (\ref{dval}).

\vspace{0.5cm}
We have checked the validity of this conjecture numerically up to $M=100$.

\paragraph{Why the expansion functions are polynomials.}\label{expfuncpol}

Conjecture 1 also lies at the heart of the argument that the functions $\mathrm{Polyn}_{k-3}$ in the expansion (\ref{Gexpgen})-(\ref{Gexpgen2}) are indeed polynomials. To prove this, we insert the expansion (\ref{Gexpgen})-(\ref{Gexpgen2}) in the equation (\ref{Ricciflat}) or (\ref{master}), assuming for the moment that $\mathrm{Polyn}_{k-3}$ are arbitrary functions, not necessarily polynomials. Expanding the equation in powers of $1\over \nu$, we obtain a series of equations of the same form as (\ref{DMop}), i.e. $D_M \mathrm{Polyn}_{M-3}=\mathrm{r.h.s.}$, where the r.h.s. depends on lower orders of perturbation theory, i.e. on $\mathrm{Polyn}_{k-3}$ with $k<M$. Next we assume that the functions $\mathrm{Polyn}_{k-3}$ with $k<M$ entering the r.h.s. are polynomials of the corresponding degree. Analyzing the terms in the r.h.s. term-by-term, one proves that the r.h.s. is a polynomial, whose degree does not exceed $M-3$. Therefore, substituting an ansatz $\mathrm{Polyn}_{M-3}=\sum\limits_{s=0}^{M-3}\,a_s \,\xi^s$ into the equation and equating the coefficients of the resulting polynomials of degree $M-3$ (note that the term in the l.h.s. of highest degree $M-2$ nicely cancels out), we obtain a set of $M-2$ linear equations in the $M-2$ variables $a_0, \ldots, a_{M-3}$: $\sum\limits_{j=0}^{M-3}\,V_{ij}\,a_j=p_i$. The nondegeneracy of the matrix $V$ is precisely equivalent to the absence of polynomial solutions of the \emph{homogeneous} equation $D_M \Pi_M=0$. Hence, if Conjecture 1 holds, one obtains a solution for the coefficients $a_i$ in terms of $p_j$. The fact that the solutions (\ref{P34}) exist for $M=3, 4$ is checked directly. $\blacksquare$

\vspace{0.5cm}
Interestingly, the conjecture may be reduced to a certain `matrix model-like' statement about the roots $x_i$ of the would-be polynomial solutions $\Pi_M$ \cite{Szego}. Indeed, suppose $\Pi_M(\xi)$ -- a polynomial of degree $M-3$ -- is a solution of
\bea\label{poleq}
\frac{d}{d\xi}\left( Q(\xi) \frac{d \Pi_M}{d\xi}\right)-\left((M-2)^2-1\right)\,(\xi-u)\, \Pi_M=0
\eea
Note that we have introduced a spectral parameter, which in our case is equal to zero: $u=0$. It is a theorem of Heine (see, for instance, \cite{Szego}) that in general there are $M-2$ values of the parameter $u$, for which there exists a polynomial solution of (\ref{poleq}).

Suppose $\Pi_M(\xi)$ has roots $x_1, \ldots, x_{M-3}$, and we can write it as $\Pi_M(\xi)=\prod\limits_{k=1}^{M-3}\;(\xi-x_k)$. Setting in (\ref{poleq}) $\xi=x_i$, we obtain the following equation:
\bea\label{matrixmodel}
\sum\limits_{j=1, j\neq i}^{M-3}\;\frac{1}{x_i-x_j}+\frac{Q'(x_i)}{2\, Q(x_i)}=0\,\quad i=1, \ldots, M-3
\eea
Setting now in (\ref{poleq}) $\xi=u$ we obtain:
\bea
0=\sum\limits_{i\neq j}\;\frac{1}{(x_i-u)(x_j-u)}=\left( \sum\limits_{i=1}^{M-3} \,\frac{1}{x_i-u}\right)^2-\sum\limits_{i=1}^{M-3}\,\frac{1}{(x_i-u)^2}
\eea
Hence, to check that $u=0$ is \emph{not} an eigenvalue, we would need to find all the solutions of the equations (\ref{matrixmodel}) and then check that $\left( \sum\limits_{i=1}^{M-3} \,\frac{1}{x_i}\right)^2-\sum\limits_{i=1}^{M-3}\,\frac{1}{x_i^2}\neq 0$ for each of them.

\vspace{0.3cm}
\textbf{Proposition 1.} The homogeneous Heun equation $D_M \Pi_M=0$ has no \underline{\emph{non-polynomial}} solutions, which are analytic at the two singular points $\xi=\xi_1$, $\xi=\xi_2$ for $M\geq 5$ and $d$ given by (\ref{dval}).

\vspace{0.5cm}
The method of solving the eigenvalue problem for the Heun equation using an expansion in hypergeometric (Jacobi) polynomials goes back to Svartholm \cite{Svartholm} (see also \cite{Slavyanov} as a general reference on Heun's equations). In our case, since the exponents of the corresponding singular points are zero, the Jacobi polynomials reduce to Legendre polynomials.

In order to make a more canonical `centering' of the Heun equation we make a change of variables
\bea
\xi \to \frac{\xi_1+\xi_2}{2}-\frac{\xi_2-\xi_1}{2}\, \xi,
\eea
bringing the equation to the canonical form
\bea\label{Heun}
\frac{d}{d\xi} \left( (1-\xi^2) (\xi-t) \frac{d P_M}{d \xi}\right)-\left((M-2)^2-1 \right) (h-\xi) P_M=0
\eea
with
\bea\label{thpar}
t=\frac{\xi_1+\xi_2-2\,\xi_0}{\xi_2-\xi_1}\quad \textrm{and}\quad h=\frac{\xi_2+\xi_1}{\xi_2-\xi_1}
\eea
We expand $P_M$ in the Legendre polynomials
\bea\label{Legendre}
P_M=\sum\limits_{k=0}^{\infty}\;a_k \,L_k(\xi).
\eea
For a function $P_M(\xi)$, \emph{analytic on the closed segment} $\xi\in[-1, 1]$, the expansion (\ref{Legendre}) is convergent in an ellipse having $\pm 1$ as its foci (\cite{Szego}, p. 245; \cite{Whittaker}, p. 322). Note that the shape of the ellipse depends on the nearest singularities of $P_M(\xi)$.

Substituting the expansion (\ref{Legendre}) in the equation (\ref{Heun}), obtain the recurrence relation
\bea\label{recursion}
h_k\;a_{k+1}-f_k\; a_k+j_k\;a_{k-1}=0
\eea
with
\bear\label{hk}
&&h_k=\frac{(k+1)\left( (k+1)^2-(M-2)^2\right)}{2k+3}\\ \label{fk}
&& f_k = t\,k(k+1)+h\,\left(1-(M-2)^2\right)\\ \label{jk}
&& j_k= \frac{k\left(k^2-(M-2)^2\right)}{2k-1}
\eear
The first thing to observe about this recurrence relation is that the values of $a_k$ for $k<M-2$ and for $k\geq M-2$ are completely independent. Suppose the recurrence relation has a nontrivial solution for $k<M-2$, i.e. that a sequence $\{a_k\}$ satisfying (\ref{recursion}) has $a_k\neq 0$ for at least one $k<M-2$. In this case $\Pi_M=\sum\limits_{k=0}^{M-3}\;a_k \,L_k(\xi)$ is a polynomial solution of the equation $D_M P_M=0$. This implies that a second solution $\tilde{\Pi}_M$, regular at the two singular points, cannot exist. Indeed, evaluating the equation at one of the singular points, say $\xi=\xi_1$, we see that for both solutions $\Pi_M'(\xi_1)=\kappa \,\Pi_M(\xi_1), \,\tilde{\Pi}_M'(\xi_1)=\kappa \,\tilde{\Pi}_M(\xi_1)$ with the same proportionality constant $\kappa$.
This implies that the Wronskian of the two solutions vanishes at $\xi=\xi_1$. On the other hand, the Wronskian of the two solutions of a second-order ODE is a constant, therefore it vanishes everywhere, hence the solutions are linearly dependent.

In any case, we have conjectured above that a polynomial solution does not exist. Hence we have to set $a_i=0, \,i=0,\;\ldots,\; M-3$, and the recursion effectively starts at $a_{M-2}$. Introducing the new variable $\tau_k=\frac{a_{k-1}}{a_k}$, we can rewrite the recurrence relation (\ref{recursion}) as follows:
\bea\label{rec}
\frac{h_k}{\tau_{k+1}}+j_k \tau_k-f_k=0
\eea
and take
\bea
\tau_{M-2}=0
\eea
as the initial condition for our recursion.

It is easy to solve the recurrence relation in the limit $k \to \infty$. Indeed, in this case we obtain a quadratic equation for $\tau_{\infty}$:
\bea
\tau_\infty^2-2t \tau_\infty+1=0,
\eea
which has the solutions
\bea
(\tau_\infty)_\pm=t\pm \sqrt{t^2-1}
\eea
The solution of the recurrence relation (\ref{recursion}) therefore behaves at large $k$ as
\bea
a_k\sim s_+ \left(\frac{1}{(\tau_\infty)_+}\right)^k+s_- \left( \frac{1}{(\tau_\infty)_-}\right)^k
\eea
It is easy to check, using (\ref{thpar}), that $t>1$, therefore $(\tau_\infty)_-<1$ and $(\tau_\infty)_+>1$. Looking back at the expansion (\ref{Legendre}), and taking into account that $L_k(1)=1, L_k(-1)=(-1)^k$, we see that the requirement of regularity of the function $P_M$ at the points $\xi=0, 1$ is equivalent to the condition $s_-=0$. We will prove below that this is not so, i.e. that the solution in fact grows as $a_k \sim \left( \frac{1}{(\tau_\infty)_-}\right)^k$, where $\frac{1}{(\tau_\infty)_-}>1$. The proof is by induction: assuming that $0<\tau_k<a$ for a suitable constant $a$, we will show that $0<\tau_{k+1}<a$. If one can take $a<1$, this is sufficient to prove that the sequence $\{a_k\}$ is exponentially growing. Details of the proof are described in Appendix \ref{proof1}.

\ssection{An example: the orthotoric metric}\label{orthometric}

In the previous sections we have demonstrated that there exists a Ricci-flat metric with $U(2)\times U(1)$ isometry on the complex cone over $\dP_1$ with \emph{at most} two parameters, which we termed $\alpha$ and $\beta$. There exists a closed expression for $G$, and hence for the metric, in a particular case when the parameters $\alpha$ and $\beta$ are related in a certain way --- this is the metric obtained in \cite{LuPope1}, as well as in \cite{MS} by means of the so-called `orthotoric' ansatz developed in \cite{Gauduchon}.

The dual potential for the orthotoric metric may be written as follows:
\begin{empheq}[box=\fbox]{align}
\hspace{0.5em}
\label{Gortho}
G_{\mathrm{ortho}}\!\!=\!\!\sum\limits_{i=1}^3\;\frac{(x-x_i)(y-x_i)}{1-x_i}\,\log{|x-x_i|}\!+\!\sum\limits_{i=1}^3\;\frac{(x-y_i)(y-y_i)}{1-y_i}\,\log{|y-y_i|}\!-\!3\,(x+y), 
\end{empheq}
where $x_i, y_i$ are respectively the roots of the following two cubic polynomials:
\bea
T_c(x)=x^3-{3\over 2} x^2+c,\quad\quad T_d(y)=y^3-{3\over 2} y^2+d=Q(y)
\eea
In particular, $y_i=\xi_i$ are the roots of $Q(y)$ that we encountered before. The moment maps $\mu, \nu$ are related to the auxiliary `orthotoric' variables $x, y$ by means of the following formulas:
\bea\label{change}
\mu=x\,y,\quad \nu-\nu_0= x+y-1
\eea
The potential (\ref{Gortho}), expressed in terms of $\mu, \nu$, satisfies the Ricci-flatness equation (\ref{Ricciflat}) with $a=-9$. Note that the function $G$ defined in this way is a solution of (\ref{Ricciflat}) for any value of $\nu_0$, since $\nu_0$ simply reflects the translational invariance of the equation (\ref{Ricciflat}) with respect to the shift $\nu \to \nu + \textrm{const.}$ However, we will subsequently fix it by the requirement that the `new origin' of the moment polytope (see Fig. \ref{mompol}) be located at $\nu=0$.

One can now introduce new variables $\{\nu, \xi\}$ according to (\ref{newvars}) and expand the function $G$ at $\nu \to \infty$. The requirement of the absence of singular terms at $\xi \to \xi_1, \xi_2$ determines the constants $\mu_0$ and $\nu_0$:
\bea
\mu_0= \xi_1 \xi_2\quad \mathrm{and}\quad \nu_0=1-(\xi_1+\xi_2)
\eea
Upon substitution of these values the expansion of the orthotoric potential $G$ in powers of ${1\over \nu}$ has the following form:
\bear\nonumber
G_{\mathrm{ortho}}&=&3\nu \left(\log{\nu}-1\right)-3\xi_0 \log{(\nu\,(\xi-\xi_0))}+\nu\,\sum\limits_{i=0}^2\;\frac{1}{1-\xi_i}\;(\xi-\xi_i)\,\log{(\xi-\xi_i)}+\\   &+&
\frac{9 \xi_0^2 (1-\xi_0)}{2(\xi-\xi_0)}\,\frac{1}{\nu}\;+\left( \frac{d-c}{2}+\frac{9\,\xi_0^3 (1-\xi_0)^2}{2\,(\xi-\xi_0)^2}\right)\,\frac{1}{\nu^2}+\\ \nonumber &+&\left(\frac{27\,(1-\xi_0)^3 \xi_0^4}{4\,(\xi-\xi_0)^3}+ (d-c)(\xi_0+{3\over 4}(\xi-1)) \right)\,\frac{1}{\nu^3}+\ldots
\eear
We see that this expansion has the general structure of (\ref{Gexpgen}) with $b=-3\, \xi_0$. Moreover, we can identify the parameters $\alpha, \beta$ of (\ref{P34}):
\bea\label{alphabetaortho}
\alpha=\frac{1}{2}\,\frac{d-c}{(-3 \,\xi_0)^3},\quad\quad \beta=\frac{3}{4}\,\frac{d-c}{(-3\, \xi_0)^4}
\eea
The fact that $\alpha$ and $\beta$ are related in this way means that the orthotoric metric is a \emph{special case} of a more general metric, in which the parameters $\alpha$ and $\beta$ are independent.

\begin{wrapfigure}{l}{0.4\textwidth}
  \centering
    \includegraphics[width=0.4\textwidth]{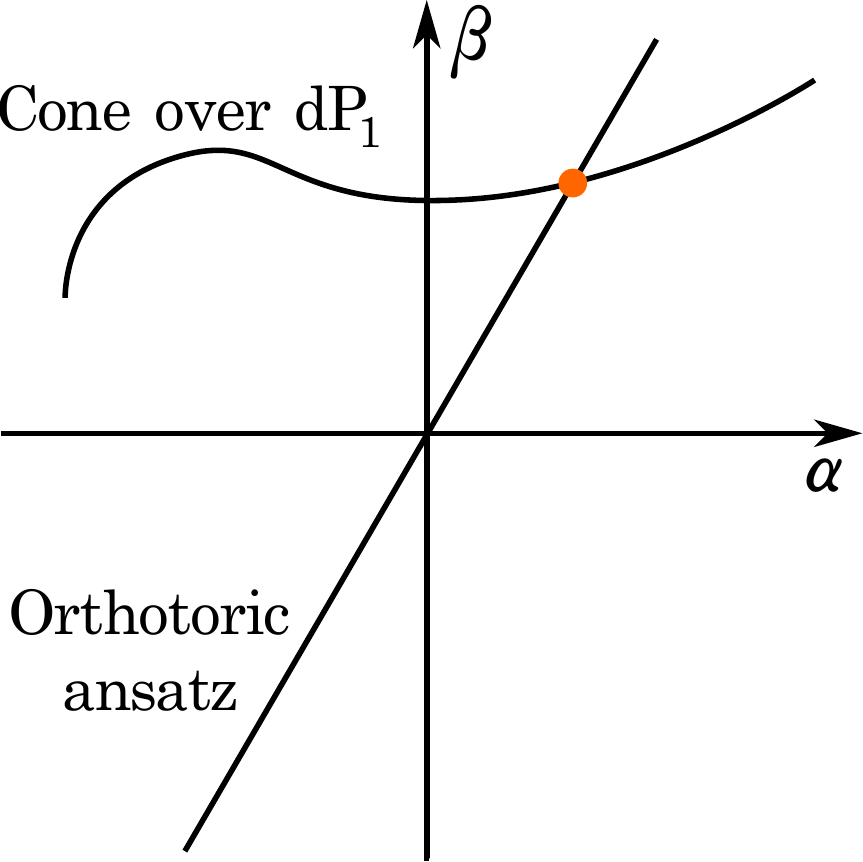}
    \caption{The red spot represents the known (orthotoric) metric on the cone over $\dP_1$.}
    \label{orthogen}
\end{wrapfigure}

It might seem from this discussion that the orthotoric potential $G_{\mathrm{ortho}}$ still possesses one nontrivial parameter $c$. However, it turns out that this parameter has to be fixed to a particular value by the requirement that the 3-rd line of the biangle in Fig. \ref{mompol} passes at a correct angle with respect to the other two lines (meaning that the topology of the manifold is indeed the one of a cone over $\dP_1$). Even in the general case, when we do not impose the orthotoric relation (\ref{alphabetaortho}) between $\alpha$ and $\beta$, we expect there to be an additional tolopogical relation between these parameters (see the discussion in \S\,\ref{conclusion} and Fig. \ref{orthogen}).

\vspace{0.2cm}
\ssubsection{Deformation of the orthotoric metric}\label{orthodeform}
\vspace{0.2cm}

To check the consistency of our conclusions regarding the parameters of the metric we wish to show directly that there is a deformation of the orthotoric metric compatible with our general considerations.

We make the following substitution for the potential $G$:
\bea
G=G_{\mathrm{ortho}}+H
\eea
Expanding the Ricci-flatness equation (\ref{Ricciflat}) around the orthotoric solution to the first order in the deformation $H$, we obtain the following remarkably simple linear equation:
\bea\label{Hlin}
\frac{1}{x}\,\frac{\dd}{\dd x} \left( T_{c}(x)\,\frac{\dd H}{\dd x}\right)-\frac{1}{y}\,\frac{\dd}{\dd y} \left( T_d(y)\,\frac{\dd H}{\dd y}\right)=0
\eea
As we discussed above, the new deformation parameter (called $\beta$) should arise in the order $\nu^{-3}$ and, since for $\nu \to \infty$ we have $x=\nu+\ldots$ (see (\ref{change})), we look for a solution of (\ref{Hlin}) with the asymptotic behavior
\bea
H=\beta\, \left( {q(y)\over x^3} +\ldots\right)\quad\quad \mathrm{as}\quad x\to\infty
\eea
In fact, one finds out that a perfectly consistent ansatz is simply requiring that $H$ is linear in $y$ to all orders in $x$, i.e.
\bea
H=\beta \, (y-1)\,f(x)
\eea
which leads to the equation for $f(x)$
\bea
\frac{d}{dx}\left( T_c(x)\,\frac{df}{dx}\right)-3\,x\, f(x)=0
\eea
Notice that this is the same equation $D_4 f=0$ encountered above, where $D_4$ is defined in (\ref{DMop}). We already know one solution $f(x)=x-1$. However, this is not the solution decaying as ${1\over x^3}$ at $x\to \infty$. The second solution is easily recovered:
\bea
f(x)=(x-1)\,\int\limits^x\,\frac{dz}{(z-1)^2\,T_c(z)}
\eea
One momentarily checks that it has the right behavior at $x\to \infty$. As a result, we have obtained the following first-order deformation:
\bea
H=\beta \, (y-1)\,f(x)=\beta \, (x-1)(y-1)\,\int\limits^x\,\frac{dz}{(z-1)^2\,T_c(z)}\;.
\eea

\ssection{Conclusion and outlook}\label{conclusion}

In this paper we have analyzed the parameter space of Ricci-flat metrics on the complex cone over a del Pezzo surface of rank one (sometimes also called the Hirzebruch surface $F_1$). In particular, using an expansion at infinity, we have found one potential new parameter $\beta$ (see (\ref{betapar})). As we discussed at the end of \S\,\ref{orthometric}, it is an additional topological requirement that this parameter should preserve the angle of the 3-rd line in Fig. \ref{mompol}. This can only be checked if the perturbation theory in $1\over \nu$ is summed.
In general we conjecture that there is a particular relation between $\beta$ and $\alpha$ that preserves the correct topology, i.e. $\beta=\beta(\alpha)$ (see Fig. \ref{orthogen}). In this case the remaining parameter is related to the size of the blown-up $\CP^1$ in the base of the cone, i.e. in the del Pezzo surface. In the metric (\ref{metric}) $\mu$ is the coefficient in front of $g_{\CP^1}$, therefore from the point of view of the moment biangle of Fig. \ref{mompol}, the size of the blown-up $\CP^1$ is measured by the $\mu$-coordinate of its lower corner, if we assume that the $\mu$-coordinate of the upper corner is fixed (we can fix it by rescaling the variables $\mu, \nu$).

Analyzing Heun's equation (\ref{DMop}), we have also proven, up to the validity of Conjecture 1 (which was checked numerically up to $M=100$), that there can be no further parameters in the metric.

In the present paper we have not analyzed the convergence of the $1\over \nu$ expansion (\ref{Gexpgen}). In the orthotoric case (\ref{Gortho}) the corresponding expansion has a finite radius of convergence. We expect that it will remain so even after the metric is deformed by the new parameter. It would be very interesting to obtain an exact formula, like (\ref{Gortho}), for the solution with two generic values of the parameters $\alpha, \beta$, and this would certainly shed light on these questions.

\vspace{0.3cm}
\textbf{Acknowledgements.}
{\footnotesize
I would like to thank Dmitri Ageev for a collaboration at an initial stage of this project, for many useful conversations and for helpful suggestions on the manuscript. I am grateful to Sergey Frolov for discussions. I am indebted to Prof.~A.A.Slavnov and to my parents for constant support and encouragement. My work was supported in part by grants RFBR 14-01-00695-a, 13-01-12405 ofi-m2 and the grant MK-2510.2014.1 of the President of Russia Grant Council.
}
\vspace{-0.3cm}
\begin{center}
\line(1,0){450}
\end{center}

\appendix
\begin{center}
{\normalfont\scshape \large \underline{Appendix}}
\end{center}

\section{Determining the physical roots $\xi_1, \xi_2$ of $Q(\xi)=0$} \label{solxi}

We showed in \S\, \ref{metricinf} that the normal bundles of the spheres embedded in the cone require that
\bea
-\frac{\xi_2}{1-\xi_2}=\frac{3 \xi_1}{1-\xi_1},
\eea
where $\xi_1$ and $\xi_2$ are \emph{both} roots of the polynomial $Q(\xi)$. This means that
\bear
&\xi_1+\xi_2+\xi_0={3\over 2},&\\
&\xi_1\xi_2+\xi_1\xi_0+\xi_2\xi_0=0,&\\
&\xi_2=\frac{3 \xi_1}{4 \xi_1-1}&
\eear
Eliminating the variables $\xi_0$ and $\xi_2$ we arrive at a cubic equation for $\xi_1$, which, however, factorizes:
\bea
(\xi_1-1)(16 \xi_1^2-4 \xi_1-3)=0
\eea
As we mentioned in \S \, \ref{leading}, the case $\xi_1 = 1$ corresponds to the case when the physical region shrinks to zero (i.e. the lines $1, 2$ in Fig. \ref{mompol} merge), so we assume that $\xi_1 \neq 1$. Then we have the two solutions:
\bear
\xi_1^{(1)}={1\over 8} (1+\sqrt{13}),\quad \xi_2^{(1)}={1\over 8} (7+\sqrt{13})\\
\xi_1^{(2)}={1\over 8} (1-\sqrt{13}),\quad \xi_2^{(2)}={1\over 8} (7-\sqrt{13})
\eear
Since $P_0''=-\frac{3 \xi}{Q(\xi)}$, in order for the metric at infinity (\ref{infmetr}) to be positive-definite, we ought to determine in which of these segments $(\xi_1^{(i)}, \xi_2^{(i)})$ the function $\frac{\xi}{Q(\xi)}$ is negative (in the whole segment). An elementary check shows that this is so only for the first segment, $(\xi_1^{(1)}, \xi_2^{(1)})$. This leads to the following value of $d$:
\bea
d=\frac{16+\sqrt{13}}{64}
\eea

\section{Perturbation theory up to $P_8(\xi)$}\label{perturbfunc}

In this appendix we summarize the results of our perturbation theory calculations for $G$ up to the 9-th order of perturbation theory, i.e. up to the function $P_8(\xi)$. The calculations were mainly carried out in $Mathematica$ and, in principle, could be extended to higher orders, subject to greater machine time. We build both the expansions in powers of $1\over \nu$, as well as in powers of $1\over x$.  These two are, of course, expressible in terms of each other, however the latter one is useful for probing the applicability of the orthotoric variables in the general case. 

\subsubsection{Rational parametrization for the space of polynomials $x^3-{3\over 2} x^2+d$}

In most calculations one encounters the roots $\xi_i$ of the polynomials of the form
\bea\label{poltyp}
Q(x)=x^3-{3\over 2} x^2+d
\eea
These can be written out explicitly in terms of Cardano's formula, however this expression is rather complicated. A better approach is to use a \emph{rational} parametrization for the space of polynomials of the form (\ref{poltyp}). Indeed, denoting the roots of such a polynomial by $\xi_0, \xi_1, \xi_2$ (as we did in the body of the paper), polynomials of the type (\ref{poltyp}) are defined by the following relations:
\bea
\xi_0+\xi_1+\xi_2={3\over 2},\quad\quad \xi_0\xi_1+\xi_0 \xi_2+\xi_1\xi_2=0
\eea
Reparametrizing the roots as $\xi_1 = \lambda_1 \xi_0,\;\xi_2=\lambda_2 \xi_0$, we arrive at a simple equation $(\lambda_1+1)(\lambda_2+1)=1$, which can be `solved' as follows: $\lambda_1+1=u,\; \lambda_2+1={1\over u}$, where $u$ is a new variable. In terms of this variable the roots are parametrized as
\bea
\xi_0={3\over 2}\,\frac{1}{u+{1\over u}-1},\quad \xi_1={3\over 2}\,\frac{u-1}{u+{1\over u}-1},\quad \xi_2={3\over 2}\,\frac{{1\over u}-1}{u+{1\over u}-1},
\eea
whereas the parameter $d$ of the polynomial $Q(x)$ is expressed as
\bea
d={27 \over 8}\,\frac{(u-1)^2}{u}\,\frac{1}{\left( u+{1\over u}-1\right)^3}
\eea

\subsection{The $(\nu, \xi)$ variables}

The expansion of the potential $G$ at $\nu \to \infty$ has the general form
\bear\label{expapp}
G=3 \nu (\log{\nu}-1)+\nu\,P_0(\xi)+b \log{(\nu\,(\xi-\xi_0))}+\sum\limits_{k=1}^\infty\;\nu^{-k}\;P_{k+1}(\xi)\\
\mathrm{with}\quad P_k(\xi)=b^k \left( \frac{(-1)^k}{k(k-1)}\,\left(\frac{1-\xi_0}{\xi-\xi_0}\right)^{k-1}+\mathrm{Polyn}_{k-3}(\xi)\right),\quad k\geq 2
\eear
and
$$
\mathrm{Polyn}_0(\xi)=\alpha,\quad \mathrm{Polyn}_1(\xi)=-\frac{2 \alpha}{3}+\beta\,(\xi-1)$$
\bear \nonumber
\mathrm{Polyn}_2(\xi)&=&\left(\frac{\alpha }{24 \xi _0^2}+\frac{\alpha }{3}-\frac{\beta  \xi _0}{3}+\frac{\beta }{6 \xi
   _0}+\frac{\beta }{6}\right)+\\ \nonumber &+& (\xi -1) \left(-\frac{\alpha }{15 \xi _0^2}-\frac{4 \beta }{15 \xi
   _0}-\frac{4 \beta }{3}\right)-(\xi -1)^2\,\frac{\alpha +4 \beta  \xi _0}{3 \xi
   _0^2}
\eear
   \bear
\nonumber
   \mathrm{Polyn}_3(\xi)&=&\frac{54 \alpha ^2 \xi _0^3-22 \alpha  \xi _0^3+6 \alpha  \xi _0^2-9 \alpha  \xi _0-2 \alpha +60
   \beta  \xi _0^4-18 \beta  \xi _0^3-36 \beta  \xi _0^2-6 \beta  \xi _0}{135 \xi _0^3}+\\ \nonumber
   &+&(\xi -1)
   \left(\frac{2 \alpha }{9 \xi _0}-\frac{\alpha }{9 \xi _0^3}-\frac{\beta }{3 \xi _0^2}+2 \beta
   \right)+(\xi -1)^2\,\frac{2  \left(3 \alpha  \xi _0+\alpha +12 \beta  \xi _0^2+3 \beta  \xi
   _0\right)}{9 \xi _0^3}+\\ \nonumber &+&(\xi -1)^3\,\frac{14  \left(\alpha +3 \beta  \xi _0\right)}{27 \xi
   _0^3}
   \eear
   \bear
    \nonumber
   \mathrm{Polyn}_4(\xi) &=& \frac{1}{1296 \xi _0^4}\,\left(-864 \alpha ^2 \xi _0^4+64 \alpha  \xi _0^4-48 \alpha  \xi _0^3+120 \alpha  \xi _0^2+8 \alpha 
   \xi _0-9 \alpha -672 \beta  \xi _0^5+\right. \\  \nonumber &+& \left.   240 \beta  \xi _0^4+432 \beta  \xi _0^3+24 \beta  \xi
   _0^2-24 \beta  \xi _0 \right)+\\ \nonumber  &+&(\xi -1)\,\frac{1}{378 \xi _0^4}\, \left(432 \alpha  \beta  \xi _0^4-164 \alpha 
   \xi _0^3-20 \alpha  \xi _0^2+112 \alpha  \xi _0+27 \alpha -960 \beta  \xi _0^4+ \right.\\ \nonumber &+& \left. 12 \beta  \xi
   _0^3+336 \beta  \xi _0^2+72 \beta  \xi _0\right)+\\ \nonumber &+&(\xi -1)^2\,\frac{1}{54 \xi _0^4}\, \left(-70 \alpha 
   \xi _0^2-14 \alpha  \xi _0+9 \alpha -252 \beta  \xi _0^3-42 \beta  \xi _0^2+24 \beta  \xi
   _0\right)-\\ \nonumber &-&(\xi -1)^3\,\frac{4  \left(28 \alpha  \xi _0+9 \alpha +84 \beta  \xi _0^2+24
   \beta  \xi _0\right)}{81 \xi _0^4}-(\xi -1)^4\,\frac{2  \left(3 \alpha +8 \beta  \xi _0\right)}{9
   \xi _0^4}
   \eear
   
The polynomial $\mathrm{Polyn}_5(\xi)$ is too complicated to be written out in full form. It turns out, however, that its significance lies in the fact that $P_8(\xi)$, or $\mathrm{Polyn}_5(\xi)$, is the first order where terms quadratic in the \emph{new} deformation parameter $\beta$ enter. For convenience we will expand this polynomial around the orthotoric point $\beta=-{\alpha\over 2 \xi_0}$ (see \ref{alphabetaortho}), i.e.
\bea
\label{polyn5}
\mathrm{Polyn}_5(\xi)=A + B \,(\beta+{\alpha\over 2 \xi_0})+C (\beta+{\alpha\over 2 \xi_0})^2,
\eea
and write out the coefficient of the quadratic term:
\bea\label{C}
C=\frac{2 \left(1232 d \xi ^2-2814 d \xi +1854 d+1232 \xi ^5-5040 \xi ^4+7560 \xi ^3-6804 \xi ^2+5103 \xi
   -2187\right)}{21 (44 d-243)}
\eea

Note that the singular denominator $44d-243$ is absent in the coefficients $A, B$. As is clear from (\ref{polyn5}), this singularity also disappears at the orthotoric point $\beta=-{\alpha\over 2 \xi_0}$. The explanation of the appearance of this singularity is that, when $d={243\over 44}$, the homogeneous Heun equation $D_8 \Pi_8=0$ has a polynomial solution. Since in our case $d\neq {243\over 44}$, this singularity does not disturb us. 
 
\subsection{The orthotoric $(x, y)$ variables}

In this appendix we set the irrelevant parameter $b=-3\xi_0$ and write out the expansion of the potential $G$ at $x\to\infty$ with fixed $y$, where $(x, y)$ are the orthotoric variables related to $(\mu, \nu)$ as follows:
\bea
\mu = x\,y,\quad \nu=x+y-\xi_1-\xi_2
\eea
In other words, we express $\nu$ and $\xi$ in terms of $x$ and $y$, expand them at large $x$ and substitute these expansions in (\ref{expapp}), obtaining a similar expansion:
\bear
&&G=3\,(x+y-{3\over 2})\,\log{x}+(x+y-\xi_1-\xi_2)\,\log{\left(-{a\over 9}\right)}-3\,x+\\ \nonumber &&\quad\quad+\,\sum\limits_{i=0}^2\,\frac{(x-\xi_i)(y-\xi_i)}{1-\xi_i}\,\log{(y-\xi_i)}\,+\sum\limits_{k=1}^\infty\;x^{-k}\,S_{k+1}(y)
\eear
The coefficient functions have a slightly simpler look in these variables:
\bear
&&S_2(y)=\frac{9}{8} (3-2 y),\\ \nonumber
&&S_3(y)=-\frac{81}{2} \alpha\,  \xi _0^2+27 \alpha\,  d-\frac{d}{2}+\frac{27}{16}-\frac{9\, y}{8}\\ \nonumber
&& S_4(y)=\xi _0^3 \left(-81 \alpha +\left(54 \alpha -\frac{1}{4}\right) y+\frac{3}{4}\right)+81 \beta  \xi _0^4
   (y-1)+\frac{3}{8} \xi _0^2 (y-3)-\frac{27}{64} (2 y-3)\\ \nonumber
   && S_5(y)=\frac{9}{640} \left(8 \xi _0^3 (54 \alpha  (16 y-21)-4 y+9)+10368 \beta  \xi _0^4 (y-1)+ \right. \\ \nonumber && \left.\quad\quad\quad+12 \xi _0^2 (4
   y-9)-54 y+81\right)
   \eear
   \bear
    \nonumber
   && S_6(y)=\frac{1}{640} \left(64 (1-54 \alpha )^2 \xi _0^6+192 (54 \alpha -1) \xi _0^5+432 \xi _0^3 (54 \alpha 
   (4 y-5)-y+2)+\right. \\ \nonumber &&\left. \quad\quad\quad +144 \xi _0^4 (972 \beta  (y-1)+1)+648 \xi _0^2 (y-2)-486 y+729\right)\\ \nonumber
   && S_7(y)=\frac{1}{1792} \left(-82944 (54 \alpha -1) \beta  \xi _0^7 (y-1)-64 \xi _0^6 ((54 \alpha -1) (378 \alpha  (2 y-3)-2
   y+9)+\right. \\ \nonumber && \left. \quad\quad\quad+1944 \beta  (y-1))-192 \xi _0^5 (54 \alpha  (8 y-15)-2 y+9)+ \right.\\ \nonumber  && \left. \quad\quad\quad+216 \xi _0^3 (54 \alpha  (32
   y-39)-8 y+15)+144 \xi _0^4 (3888 \beta  (y-1)-2 y+9)\right. \\ \nonumber && \left. \quad\quad\quad +324 \xi _0^2 (8 y-15)-729 (2 y-3) \right)
\eear
The function $S_8(y)$ is, once again, too complicated to write out, but, just as above, we will write out the coefficient of $(\beta+{\alpha\over 2 \xi_0})^2$ in its expansion around the orthotoric point. More precisely, if $S_8(y)=A_1+B_1 (\beta+{\alpha\over 2 \xi_0})+C_1 (\beta+{\alpha\over 2 \xi_0})^2$, then
\bear
C_1&=& \frac{2\, (-3 \xi_0)^8}{21 (44 d-243)} \,\left(2 d \left(616 y^2-1407 y+927\right)+1232 y^5-5040 y^4+ \right.\\ \nonumber && \left.+7560 y^3-6804 y^2+5103
   y-2187 \right)
\eear
This is essentially the same polynomial as $C$ above (\ref{C}). Note that while $A_1$ and $B_1$ are linear in $y$, just like all the previous functions $S_2\,\ldots\, S_7$, $C_1$ is a polynomial of degree 5. In higher orders of perturbation theory the degree of the polynomial will grow accordingly. This means that the orthotoric variables are not well-suited for the description of the metric in its most general form.

\section{To the proof of Proposition 1.} \label{proof1}

The key technical inequality that we will need to prove is as follows:
\bea\label{masterineq}
f_k-a j_k-{1\over a} h_k >0\quad \textrm{for all}\quad k>M-2\quad{\textrm{and some}\;\; a:}\quad 0<a<1\, ,
\eea
where $f_k, j_k, h_k$ have been defined in (\ref{hk})-(\ref{jk}). Once we have proven this inequality, suppose $0<\tau_k<a$. Then
\bea\label{reck1}
\tau_{k+1}=\frac{h_k}{f_k-j_k \tau_k}>0,
\eea
since $j_k>0$ and $h_k>0$ for $k>M-2$, and it follows from (\ref{masterineq}) that ${f_k\over j_k}>a>\tau_k$. Besides, since, according to (\ref{masterineq}), $f_k-\tau_k j_k> f_k - a j_k> {1\over a} h_k$, (\ref{reck1}) implies
\bea
\tau_{k+1}<a .
\eea
\vspace{-1.5cm}
\begin{center}
\line(1,0){50}
\end{center}

In order to prove (\ref{masterineq}), first of all we make some elementary estimates:
\bear
h_k<\frac{1}{2} \left((k+1)^2-(M-2)^2 \right)\\
j_k< \left( (k+1)^2-(M-2)^2\right),
\eear
\vspace{-0.2cm}
hence
\bea
f_k-a j_k-{1\over a} h_k >  t \,k (k+1) + h\, (1-(M-2)^2)-b ((k+1)^2-(M-2)^2):= \phi_k
\eea
with
\bea
b=a+{1\over 2 a}\,.
\eea
$\phi_k$ is a quadratic function in $k$, so in order to prove that $\phi_k>0$ for $k\geq M-2$ we will show that $\phi_{M-3}>0$ and $\phi'_k>0$ for $k>0$. First of all,
\bea
\phi_{M-3}=(M-3)(t-h) \left( M+\frac{h-2 t}{t-h}\right)
\eea
In the case of interest $\phi_{M-3}>0$ for $M\geq 5$. To ensure that $\phi_k$ is a growing parabola we require $t>b$ and one easily shows that for $t>2b$ the bottom of the parabola lies at $k<0$. Therefore for $t>2b$ we have $\phi_k>0$ for $k\geq M-2$, and therefore
\bea
f_k-a j_k-{1\over a} h_k >0.
\eea
Now, the requirement $t>2b$ means that
\bea
a^2-{t\over 2} a+{1\over 2}<0
\eea
This is easily satisfied for $a={1\over 2}$, since $t=\sqrt{13}>3$. Therefore we have proven that $\tau_k < {1\over 2}$, so that ${a_{k}\over a_{k-1}}>2$, which implies in particular $\underset{k\to \infty}{\lim} {a_{k}\over a_{k-1}} >1$, so that the expansion (\ref{Legendre}) is divergent at the two singular points of interest: $\xi=0$, $\xi=1$.

This completes the proof of \textbf{Proposition 1} $\blacksquare$

\vspace{-1.3cm}
\renewcommand\refname{\begin{center} \centering\normalfont\scshape  References\end{center}}
\bibliography{refsdelpezzoarxiv}

\begin{thebibliography}{10}

\bibitem{EH}
T.~Eguchi and A.~J. Hanson, ``{Asymptotically Flat Selfdual Solutions to
  Euclidean Gravity},'' {\em Phys.Lett.}, vol.~B74, p.~249, 1978.

\bibitem{GH}
G.~Gibbons and S.~Hawking, ``{Gravitational Multi - Instantons},'' {\em
  Phys.Lett.}, vol.~B78, p.~430, 1978.

\bibitem{CdO}
P.~Candelas and X.~C. de~la Ossa, ``{Comments on Conifolds},'' {\em
  Nucl.Phys.}, vol.~B342, pp.~246--268, 1990.

\bibitem{LuPope1}
W.~Chen, H.~Lu, and C.~Pope, ``{Kerr-de Sitter black holes with NUT charges},''
  {\em Nucl.Phys.}, vol.~B762, pp.~38--54, 2007.

\bibitem{Gauduchon}
V.~A. D.~M. {Calderbank} and P.~{Gauduchon}, ``{Hamiltonian 2-forms in K\"ahler
  geometry. I: General theory.},'' {\em {J. Differ. Geom.}}, vol.~73, no.~3,
  pp.~359--412, 2006.

\bibitem{Pedersen}
H.~{Pedersen} and Y.~{Poon}, ``{Hamiltonian constructions of K\"ahler-Einstein
  metrics and K\"ahler metrics of constant scalar curvature.},'' {\em {Commun.
  Math. Phys.}}, vol.~136, no.~2, pp.~309--326, 1991.

\bibitem{Guillemin}
V.~{Guillemin}, ``{Kaehler structures on toric varieties.},'' {\em {J. Differ.
  Geom.}}, vol.~40, no.~2, pp.~285--309, 1994.

\bibitem{Bykov}
D.~Bykov, ``{The K\"ahler metric of a blow-up},'' arXiv:1307.2816, 2013.

\bibitem{PZTmain}
L.~A. Pando~Zayas and A.~A. Tseytlin, ``{3-branes on spaces with $R \times S^2
  \times S^3$ topology},'' {\em Phys.Rev.}, vol.~D63, p.~086006, 2001.

\bibitem{Szego}
G.~{Szeg\"o}, ``{Orthogonal polynomials. 4th ed.}.'' {American Mathematical
  Society (AMS), 432 p.}, 1975.

\bibitem{Svartholm}
N.~{Svartholm}, ``{Die L\"osung der Fuchsschen Differentialgleichung zweiter
  Ordnung durch hypergeometrische Polynome.},'' {\em {Math. Ann.}}, vol.~116,
  pp.~413--421, 1939.

\bibitem{Slavyanov}
S.~Y. {Slavyanov} and W.~{Lay}, {\em {Special functions. A unified theory based
  on singularities. With a foreword by Alfred Seeger.}}
\newblock Oxford: Oxford University Press, 2000.

\bibitem{Whittaker}
E.~{Whittaker} and G.~{Watson}, ``{A course of modern analysis. An introduction
  to the general theory on infinite processes and of analytic functions; with
  an account of the principal transcendental functions. 4th ed., reprinted.}.''
  {Cambridge: At the University Press. 608 p. (1962).}, 1962.

\bibitem{MS}
D.~Martelli and J.~Sparks, ``{Resolutions of non-regular Ricci-flat Kahler
  cones},'' {\em J.Geom.Phys.}, vol.~59, pp.~1175--1195, 2009.

\end{thebibliography}
\bibliographystyle{ieeetr}

\end{document}